\documentclass[twocolumn,floatfix,superscriptaddress,a4paper,showpacs,showkeys,nofootinbib,reprint,prl]{revtex4-1}
\usepackage{epsfig}
\usepackage{latexsym}
\usepackage{xspace}
\usepackage[colorlinks=true,linktocpage=true,linkcolor=blue,citecolor=blue,allcolors=blue]{hyperref}
\usepackage[utf8]{inputenc}
\usepackage{indentfirst}
\usepackage{enumerate}
\usepackage{color}

\usepackage[caption=false,position=top]{subfig}

\usepackage{amsmath}
\usepackage{amssymb}
\usepackage[english]{babel}
\usepackage{url}
\newcommand{\eq}[1]{\begin{align} #1 \end{align}}

\begin{document}


\title{
Nucleosynthesis in heavy-ion collisions at the LHC via the Saha equation
}

\author{Volodymyr Vovchenko}
\affiliation{
Institut f\"ur Theoretische Physik,
Goethe Universit\"at Frankfurt, Max-von-Laue-Str. 1, D-60438 Frankfurt am Main, Germany}
\affiliation{Frankfurt Institute for Advanced Studies, Giersch Science Center, Ruth-Moufang-Str. 1, D-60438 Frankfurt am Main, Germany}

\author{Kai Gallmeister}
\affiliation{
Institut f\"ur Theoretische Physik,
Goethe Universit\"at Frankfurt, Max-von-Laue-Str. 1, D-60438 Frankfurt am Main, Germany}

\author{J\"urgen Schaffner-Bielich}
\affiliation{
Institut f\"ur Theoretische Physik,
Goethe Universit\"at Frankfurt, Max-von-Laue-Str. 1, D-60438 Frankfurt am Main, Germany}

\author{Carsten Greiner}
\affiliation{
Institut f\"ur Theoretische Physik,
Goethe Universit\"at Frankfurt, Max-von-Laue-Str. 1, D-60438 Frankfurt am Main, Germany}

\begin{abstract}
The production of light (anti-)(hyper-)nuclei in heavy-ion collisions at the LHC is considered in the framework of the Saha equation, making use of the analogy between the evolution of the early universe after the Big Bang and that of ``Little Bangs'' created in the lab.
Assuming that disintegration and regeneration reactions involving light nuclei proceed in relative chemical equilibrium after the chemical freeze-out of hadrons, their abundances are determined through the famous cosmological Saha equation of primordial nucleosynthesis and show no exponential dependence on the temperature typical for the thermal model.
A quantitative analysis, performed using the hadron resonance gas model in partial chemical equilibrium, shows agreement with experimental data of the ALICE collaboration on 
d, $^3$He, $^3_\Lambda$H, and $^4$He
yields for a very broad range of temperatures at $T  \lesssim 155$~MeV.
The presented picture is supported by the observed suppression of resonance yields in central Pb--Pb collisions at the LHC.
\end{abstract}



\maketitle


\paragraph*{Introduction.}

The yields of light (anti-)(hyper-)nuclei such as deuteron~(d), helium-3~($^3$He), hypertriton~($^3_\Lambda$H), and helium-4~($^4$He) have recently been measured in Pb-Pb collisions at the LHC by the ALICE collaboration~\cite{Adam:2015vda,Adam:2015yta,Acharya:2017bso}.
Common approaches used to describe the production of these loosely-bound objects include the thermal-statistical approach~\cite{Mekjian:1977ei,Siemens:1979dz,Hahn:1986mb,BraunMunzinger:1994iq,Andronic:2010qu} and the coalescence model~\cite{Butler:1961pr,Butler:1963pp,Csernai:1986qf,Scheibl:1998tk}, a determination of the production mechanism is of great interest~[see, e.g., Refs.~\cite{Mrowczynski:2016xqm,Sun:2017xrx,Sun:2018jhg,Zhang:2018euf,Zhao:2018lyf,Sombun:2018yqh,Bellini:2018epz,Braun-Munzinger:2018hat} for recent results within these two approaches].
The measured yields have been observed to agree remarkably well with a thermal model calculation at a temperature $T_{\rm ch} \simeq 155$~MeV of the conventional chemical freeze-out of hadrons~\cite{Becattini:2012xb,Petran:2013lja,Andronic:2017pug}, while the available transverse momentum spectra of both nuclei and stable hadrons are characterized by a lower kinetic freeze-out temperature $T_{\rm kin} \simeq 100-115$~MeV~\cite{Adam:2015vda}.
These observations suggest that certain thermal aspects are present in the production mechanism of loosely-bound objects. 
On the other hand, a survival of these fragile objects in hot and dense thermal environment at $T = T_{\rm ch}$ all the way to their detection would seem surprising, given their small binding energies relative to the system temperature~(the binding energy of deuteron and $\Lambda$ in $^3_\Lambda$H is of order 130~keV~\cite{Juric:1973zq}), and, for instance, the known large pion-deuteron break-up cross section~\cite{Garcilazo:1982yc}.
We aim to shed light on this question by making use of an analogy between the post-chemical freeze-out expansion of matter in heavy-ion collisions at the LHC and 
primordial nucleosynthesis stage in the early universe.
The analogy to the explosive big bang nucleosynthesis has been considered long time ago for intermediate energy heavy-ion collisions~\cite{Mekjian:1978zz}, and certain aspects of our approach are similar. A crucial new point here is the role of the mesonic component, which dominates at the LHC conditions and resembles photons in the early universe.

\paragraph*{Saha equation.}

The evolution of abundances in the early universe between the stage of neutron-proton ratio freeze-out~($T \sim 1$~MeV) and before the deuterium bottleneck~($T \sim 0.1$~MeV) is commonly described within the framework of nuclear statistical equilibrium. There, the nuclear formation and disintegration reactions, such as e.g. $p + n \longleftrightarrow \text{d} + \gamma$, are chemically equilibrated and the abundances are described by the nuclear equivalent of the Saha ionization equation~\cite{Kolb:1990vq}.
The entropy is carried almost exclusively by the photons, owing to a very small baryon-to-photon ratio $\eta \sim 10^{-10}$.
The resulting nuclear mass fractions read~\cite{Kolb:1990vq}
\eq{
\label{eq:BBN}
X_A & = d_A \, \left[\zeta(3)^{A-1} \, \pi^{\frac{1-A}{2}} \, 2^{\frac{3A-5}{2}}\right] \, A^{\frac{5}{2}} \, \left(\frac{T}{m_N} \right)^{\frac{3}{2}(A-1)} \nonumber \\
& \qquad \times \eta^{A-1} \, X_p^Z \, X_n^{A-Z} \, \exp\left(\frac{B_A}{T} \right)~,
}
with $d_A$, $Z$, $A$, and $B_A$ being the degeneracy factor, the electric charge, the mass number, and the binding energy of the given nucleus, respectively, $m_N \simeq 938$~MeV/$c^2$ is the nucleon mass.
The abundances are very sensitive to the value of $\eta$.

We argue that a description based on the Saha equation is relevant to describe the abundances of light nuclei produced in heavy-ion collisions at the LHC.
There, the produced meson-dominated matter is assumed to evolve in full chemical equilibrium above the chemical freeze-out temperature $T_{\rm ch}$.
The rest frame number densities of various hadron species are given by their chemical equilibrium values at $T \gtrsim T_{\rm ch}$:
\eq{
n_i^{(0)} = \frac{d_i m_i^2 T}{2 \pi^2} \, K_2(m_i/T)~.
}
Here the chemical potentials are set to zero as the produced matter is observed to be baryon-symmetric~\cite{Abelev:2013vea}.
At $T = T_{\rm ch}$ the chemical equilibrium is lost and the abundances of stable hadrons such as pions and protons are frozen.
The observation of a smaller kinetic freeze-out temperature $T_{\rm kin}$ in central collisions at LHC~\cite{Adam:2015vda}, RHIC~\cite{Adamczyk:2017iwn}, and SPS~\cite{Anticic:2016ckv}, however, suggests that (pseudo-)elastic reactions are maintained for a longer time which keep the system in kinetic equilibrium deep in the hadronic phase~\cite{Shuryak:2014zxa}. 
The hadron number densities attain fugacity factors and given at $T < T_{\rm ch}$ by $n_i = n_i^{(0)} \, e^{\mu_i/T}$.

The evidently large nuclei break-up cross sections suggest that reactions of a type $X + A \leftrightarrow X + \sum_i A_i$ proceed in relative chemical equilibrium after the chemical freeze-out of hadrons.
Here $A_i$ are components of the nucleus~(protons, neutrons, hyperons, and/or lighter nuclei), and $X$ is some other particle~(e.g. a pion).
The Saha equations, dictated by the detailed balance principle, determine the relation between the densities of nuclei and their constituents:
\eq{
\frac{n_A}{\prod_i n_{A_i}} = \frac{n_A^{(0)}}{\prod_i n_{A_i}^{(0)}},
}
which entails $\mu_A = \sum_i \mu_{A_i}$.
Explicit expressions for a number of common (hyper-)nuclei are $\mu_d = \mu_p + \mu_n$, $\mu_{^3{\rm He}} = 2 \mu_p + \mu_n$, $\mu_{_\Lambda^3{\rm H}} = \mu_p + \mu_n + \mu_\Lambda$, and $\mu_{^4{\rm He}} = 2 (\mu_p + \mu_n)$.
The yield of a (hyper-)nucleus $A$ at temperatures $T < T_{\rm ch}$ is then given by
\eq{\label{eq:NA}
N_A (T) = \frac{d_A m_A^2 T}{2 \pi^2} \, K_2(m_A/T) \, e^{\mu_A/T} \, V.
}
In order to proceed it is necessary to determine $\mu_A$ and $V$ at given temperature.

The full solution at $T<T_{\rm ch}$ would assume an isentropic expansion of a hadron resonance gas~(HRG) in partial chemical equilibrium~(PCE)~\cite{Bebie:1991ij}.
It is instructive, however, to consider first a simplified scenario in full analogy to the cosmological setup, where the results can be obtained explicitly.
First, we observe that the majority of entropy is carried by the mesonic matter. An estimate of the baryon-to-meson ratio $\eta_B$ at the LHC is given by the measured yields of protons and pions in 0-10\% central Pb--Pb collisions at the LHC. This yields $\eta_B \approx (4/3) \langle p \rangle /(\langle \pi^- \rangle + \langle \pi^+ \rangle) \simeq 0.03$, where the factor $4/3$ takes into account estimates for the unmeasured yields of neutrons and $\pi^0$.
Mesons therefore play a similar role as the photons during the evolution of the early universe -- they drive the entropy conservation during the expansion.
Assumption of effectively massless mesonic degrees of freedom yields the following condition of entropy conservation $S \approx d_M 2 \pi^2 T^3 V/45 = \text{const}$, where $d_M$ accounts for the effective degrees of freedom. One readily obtains that the volume scales as $V \sim T^{-3}$:
\eq{\label{eq:V}
\frac{V}{V_{\rm ch}} = \left( \frac{T_{\rm ch}}{T} \right)^3~.
}

The chemical potentials of baryons are obtained from the conservation of the numbers of stable baryons.
Applying the non-relativistic approximation one has
\eq{
N_i(T) \simeq d_i \, \left( \frac{m_i T}{2\pi} \right)^{\frac{3}{2}} \, e^{-m_i/T} \, e^{\mu_i/T} \, V_{\rm ch} \, \left( \frac{T_{\rm ch}}{T} \right)^3
}
for baryons such as $N$, $\Lambda$, etc.
Their number conservation requires $N_i(T) = N_i(T_{\rm ch})$ with  $\mu_i^{\rm ch} = 0$. This yields
\eq{\label{eq:muanalyt}
\mu_i \simeq \frac{3}{2} \, T \, \ln\left(\frac{T}{T_{\rm ch}} \right) + m_i \, \left(1 - \frac{T}{T_{\rm ch}}\right).
}
Inserting Eq.~\eqref{eq:muanalyt} into~\eqref{eq:NA} gives the yields of light nuclei at $T < T_{\rm ch}$:
\eq{\label{eq:drat}
\frac{N_A (T)}{N_A (T_{\rm ch})} \simeq \left( \frac{T}{T_{\rm ch}} \right)^{\frac{3}{2} (A-1)}  \, \exp\left[ B_A \left( \frac{1}{T} - \frac{1}{T_{\rm ch}} \right) \right]~.
}
The nuclei yields decrease with temperature as $T^{3(A-1)/2}$ in this simplified picture, as long as $B_A \ll T$. This is very different from the strong exponential dependence 
\eq{\label{eq:drateq}
\left[ \frac{N_A (T)}{N_A (T_{\rm ch})} \right]_{\rm eq.} \simeq \left( \frac{T}{T_{\rm ch}} \right)^{-\frac{3}{2}}  \, \exp\left[ -m_A \left( \frac{1}{T} - \frac{1}{T_{\rm ch}} \right) \right]~.
}
in the  standard chemical equilibrium thermal model approach, stemming from the large masses of nuclei, $m_A \gg T$.
Thus, a proper consideration of the strong break-up and regeneration reactions leads to an essential modification of the standard thermal model picture.

It can be even more instructive to consider the ratio of the yields of ordinary nuclei~(without strangeness) to protons.
One obtains an intriguing result
\eq{\label{eq:SahaLHC}
\frac{N_A(T)}{N_p} & = d_A \, \left[ (d_M)^{A-1} \, \zeta(3)^{A-1} \, \pi^{\frac{1-A}{2}} \, 2^{-\frac{1+A}{2}} \right] \, A^{3/2} \nonumber \\
& \quad \times \left( \frac{T}{m_N} \right)^{\frac{3}{2} (A-1)} \, \eta_B^{A-1} \, \exp\left(\frac{B_A}{T}\right).
}
Here $\eta_B \equiv N_N / N_M$ is the constant nucleon-to-meson ratio, where we assume $N_p = N_n = N_N/2$ and massless mesons, $N_M = d_M \, [\zeta(3) / \pi^2] V \, T^3$.
Similarity of this result to Eq.~\eqref{eq:BBN} for nuclei abundances in the early universe is evident.

Unfortunately, the simple result~\eqref{eq:SahaLHC} is not fully applicable for a quantitative analysis of light nuclei production at the LHC.
To obtain this result we neglected the feeddown from decays of baryonic resonances and assumed that the mesonic degrees of freedom are massless.
Moderate, yet significant corrections to these two approximations are expected at the LHC, especially regarding the feeddown.
In fact, more than half of protons at $T \simeq T_{\rm ch}$ stem from decays of baryonic resonances in a HRG model estimation.

\paragraph*{Full numerical calculation.}

Here we consider full HRG which evolves after the hadronic chemical freeze-out in a state of PCE~\cite{Bebie:1991ij}. In addition to elastic scatterings and the disintegration and regeneration reactions involving (anti-)(hyper-)nuclei, the decay and regeneration reactions like $\pi \pi \leftrightarrow \rho$, $\pi \text{K} \leftrightarrow \text{K}^*$, $\pi \text{N} \leftrightarrow \Delta$,  $\pi \text{N} \leftrightarrow \text{N}^*$, involving all strongly decaying resonances, are assumed to proceed in relative chemical equilibrium and maintain kinetic equilibrium in the system. 
This assumption is supported by the dominance of pseudo-elastic reactions with resonance formation in meson-baryon and meson-meson scatterings~\cite{Tanabashi:2018oca}. 
The effective chemical potentials $\tilde{\mu}_j$ of all species are given by
\eq{
\tilde{\mu}_j = \sum_{i \in \rm stable} \, \langle n_i \rangle_j \, \mu_i~,
}
where $\mu_i$ are the chemical potentials of species with a conserved total yield is conserved after the chemical freeze-out. 
$\langle n_i \rangle_j$ is the mean number of hadron species $i$ resulting from decays of hadron species $j$. In case $j$ is a light nucleus, $\langle n_i \rangle_j$ corresponds to the number of species $i$ in its hadron content.
The chemical potentials $\mu_i$ and the volume $V$ are determined from the conditions of PCE -- conservation of total yields of stable hadrons after the chemical freeze-out and the isentropic expansion:
\eq{
\label{eq:pce:Ni}
\sum_{j \in {\rm hrg}} \langle n_i \rangle_j \, n_j(T, \tilde{\mu}_j) \, V & = N_i^{\rm eff}(T_{\rm ch}), ~~ i \in \rm stable,  \\
\label{eq:pce:S}
\sum_{j \in {\rm hrg}} s_j(T, \tilde{\mu}_j) \, V & = S (T_{\rm ch})~.
}
Here the index $j$ runs over all hadrons, resonances and light nuclei considered.

We employ an extended version of the \texttt{Thermal-FIST} package~\cite{Vovchenko:2019pjl} in our calculations, where a numerical solver of Eqs.~\eqref{eq:pce:Ni} and~\eqref{eq:pce:S} has been additionally implemented.
We assume that yields of all hadrons stable under strong interactions are frozen after the chemical freeze-out. This includes pions, nucleons, $\eta$, $\eta'$, kaons, $\Lambda$, $\Sigma$'s, $\Xi$'s, $\Omega$ as well their antiparticles.
Effects of quantum statistics are included for all particles, while finite resonance widths and excluded volume corrections are neglected.

\begin{figure}[t]
  \centering
  \includegraphics[width=.48\textwidth]{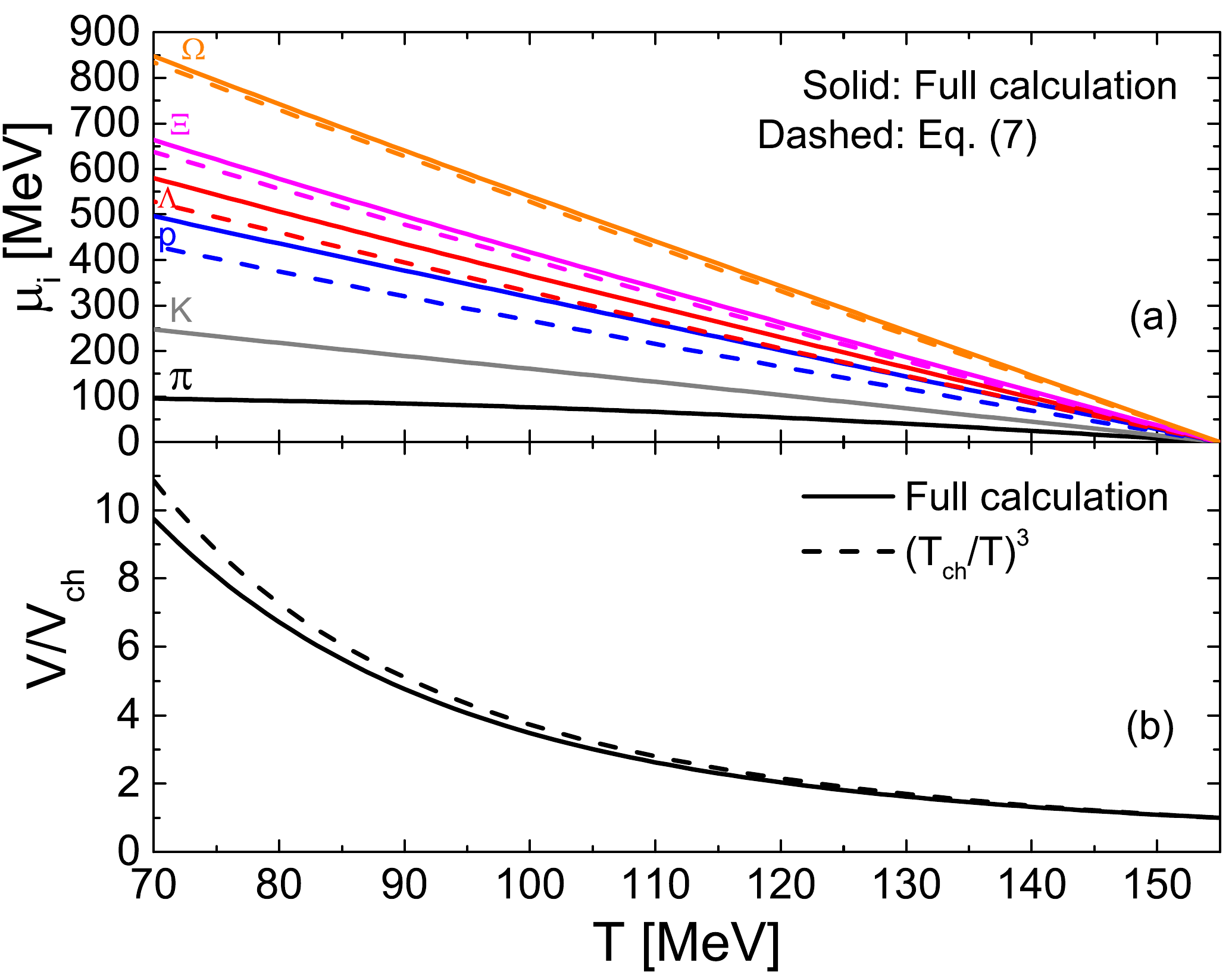}
  \caption{
   Temperature dependence of (a) effective chemical potentials of pions~(black), kaons~(grey), protons~(blue), $\Lambda$~(red), $\Xi$~(magenta), and $\Omega$~(orange), and (b) the volume ratio $V/V_{\rm ch}$, evaluated at $T<T_{\rm ch}$ within the HRG in PCE~(solid lines).
   Dashed lines in (a) depict calculations using~Eq.~\eqref{eq:muanalyt}, while the dashed line in (b) corresponds to $V/V_{\rm ch} = (T_{\rm ch}/T)^3$~[Eq.~\eqref{eq:V}].
  }
  \label{fig:muV}
\end{figure}

The chemical freeze-out conditions are obtained by fitting the measured midrapidity yields of pions, kaons, $K_0^S$, $\phi$, protons, $\Lambda$, $\Xi^-$, $\Omega$ in 0-10\% most central Pb--Pb collisions within the chemical equilibrium thermal model~(see Ref.~\cite{Vovchenko:2018fmh} for details). This yields $T_{\rm ch} = 155$~MeV, $V_{\rm ch} = 4700$~fm$^3$, and $S_{\rm ch} = 11044$. Note that both $V_{\rm ch}$ and $S_{\rm ch}$ correspond to one unit of rapidity.
Evolution at $T < T_{\rm ch}$ is described by Eqs.~\eqref{eq:pce:Ni} and~\eqref{eq:pce:S}.
The resulting temperature dependence of chemical potentials of $\pi$, K, p, $\Lambda$, $\Xi$'s, and $\Omega$ as well as of the volume is depicted in Fig.~\ref{fig:muV}.
The obtained values of the chemical potentials are rather typical for PCE HRG model applications at the LHC or RHIC~\cite{Hirano:2002ds,Kolb:2002ve}.
The approximate analytic result~[Eq.~\eqref{eq:muanalyt}] reproduces quantitatively the results for $\Xi$ and $\Omega$, but underestimates the chemical potentials of protons and $\Lambda$.
This underestimation is important for quantitative studies, as the $\mu_i$ values enter the exponent when computing the light nuclei yields,~see Eq.~\eqref{eq:NA}.
The volume, on the other hand, is reproduced fairly well by the simple relation~\eqref{eq:V}, yielding $d_M \simeq 11$-$13$ as an estimate for an effective degeneracy of ``massless'' degrees of freedom.
The baryon-to-meson ratio in full numerical calculation is $\eta_B \simeq 0.05$.

\begin{figure}[t]
  \centering
  \includegraphics[width=.49\textwidth]{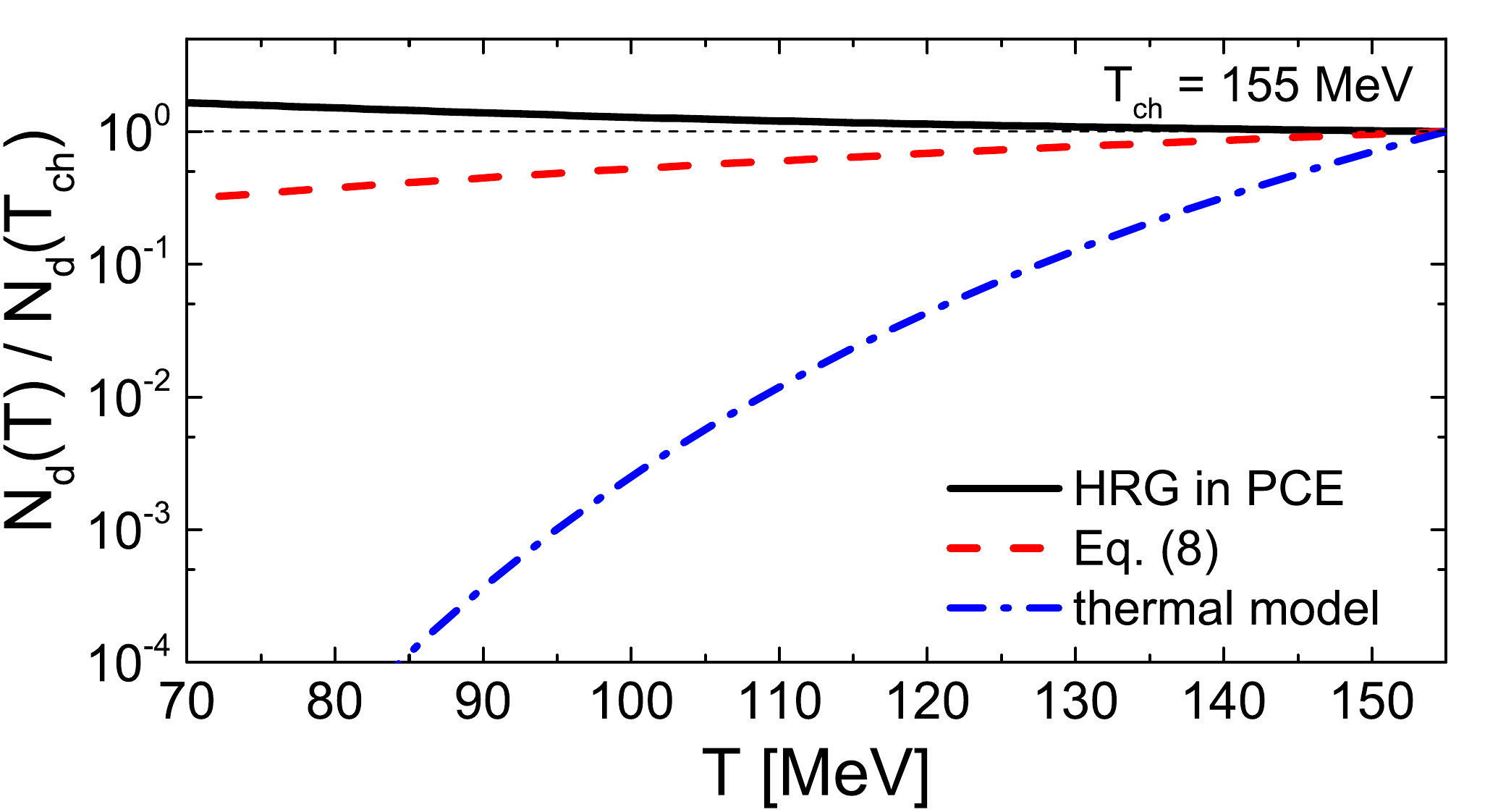}
  \caption{
   Temperature dependence of the deuteron yield relative to the one at 
   $T_{\rm ch} = 155$~MeV, calculated within the Saha equation approach using HRG in PCE~(solid black line) and simplified analytic result~(dashed red line), and within the chemical equilibrium thermal model~(dash-dotted blue line).
  }
  \label{fig:Ndrelative}
\end{figure}

Temperature dependence of the deuteron yield relative to the one at $T = T_{\rm ch}$ is depicted in Fig.~\ref{fig:Ndrelative}. The deuteron yield shows a mild increase as the temperature is lowered in the full calculation. The analytic result~\eqref{eq:drat}~[or Eq.~\eqref{eq:SahaLHC}] within the simplified approach shows instead a decrease, which is relatively mild on a logarithmic scale. 
These two results are in stark contrast to the standard chemical equilibrium thermal model, where the deuteron yield decreases with temperature exponentially.
The behavior for other (anti-)(hyper-)nuclei is qualitatively the same.

\begin{figure*}[t]
  \centering
  \includegraphics[width=.99\textwidth]{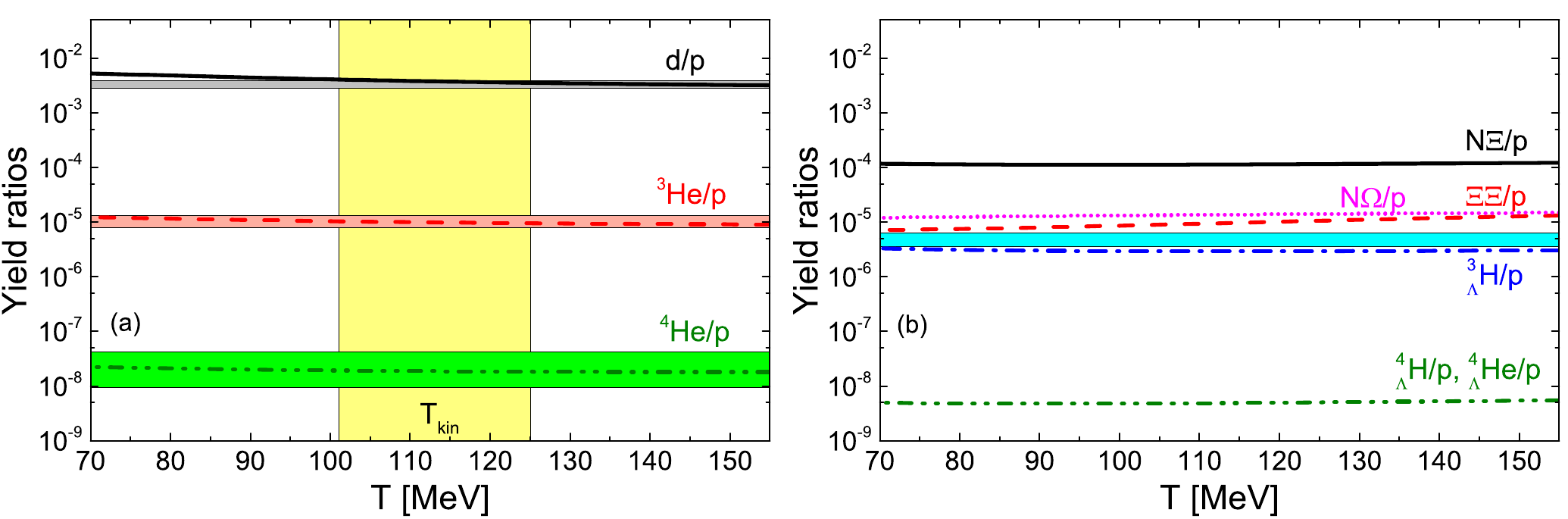}
  \caption{
   Temperature dependence of yields ratios (a) d/p~(solid black line), $^3$He/p~(dashed red line), $^4$He/p~(double-dot-dashed green line), and (b) $N\Xi$/p~(solid black line),  $N\Omega$/p~(dotted magenta line),
   $\Xi\Xi$/p~(dashed red line),
   $^3_\Lambda$H/p~(dot-dashed blue line), and $^4_\Lambda$H/p and $^4_\Lambda$He/p~(double-dot-dashed green line), evaluated at $T<T_{\rm ch}$ using the Saha equation approach and HRG in PCE.
   The horizontal bands correspond to the data of the ALICE collaboration for most central Pb--Pb collisions~\cite{Adam:2015vda,Adam:2015yta,Acharya:2017bso}.
   The data point for $^3_\Lambda$H is reconstructed assuming a 25\% branching ratio of the $^3_\Lambda\text{H} \to ^3 \text{He} + \pi$ decay~\cite{Adam:2015yta}.
   The vertical yellow band in (a) corresponds to the kinetic freeze-out temperature $T_{\rm kin} = 113 \pm 12$~MeV extracted from blast-wave fits to the transverse momentum spectra of $\pi$, $\text{K}$, protons, d, and $^3$He~\cite{Adam:2015vda}.
  }
  \label{fig:nuclei-vs-T}
\end{figure*}

Figure~\ref{fig:nuclei-vs-T} presents the temperature dependence of ratios of various light (hyper-)nuclei to the yields of protons at $T < T_{\rm ch}$.
These include d, $^3$He, $^4$He, $^3_\Lambda$H, $^4_\Lambda$H, $^4_\Lambda$He as well as hypothetical two-baryon bound states N$\Xi$, N$\Omega$, and $\Xi\Xi$.
An experimental search for the latter three~(as well as for $^4_\Lambda$H and $^4_\Lambda$He) is planned in Runs 3 and 4 at the LHC~\cite{Citron:2018lsq}.
As we work here with net baryon free matter, the results for anti-nuclei are identical.
All (hyper-)nuclei yields considered show
a mild temperature dependence after the chemical freeze-out, and agree quite well with experimental data where available.
The deuteron yield does show a notable increase at lower temperatures, indicating that an isentropic expansion after the chemical freeze-out does \emph{not} necessarily imply a fixed d/p ratio at the LHC energies, in contrast to expectations for lower collision energies~\cite{Siemens:1979dz}.
Nevertheless, the temperature dependence of the d/p ratio is still relatively mild. It is found to be sensitive to the number of baryonic resonances included in an HRG. In the extreme case where all baryonic resonances are excluded from the HRG particle list, the d/p ratio in fact shows a \emph{decrease} as one goes to lower temperatures, the dashed line in Fig.~\ref{fig:Ndrelative} represents such a behavior.
The agreement with the data for d/p, $^3$He/p, $^3_\Lambda$H/p and $^4$He/p is good for a very broad range of temperatures below $T_{\rm ch}$.
Our predictions suggest that the yields of N$\Xi$ and N$\Omega$ change very little at $T < T_{\rm ch}$, while the yield of $\Xi\Xi$ might be suppressed by up to a factor two relative to the standard thermal model prediction.

An observation of approximately identical light nuclei yield ratios evaluated at chemical and kinetic freeze-outs using effective chemical potentials for the hadronic phase was recently pointed out in Ref.~\cite{Xu:2018jff}. Our results provide a natural explanation for this phenomenon in terms of the Saha equation treatment of the break-up and regeneration reactions $X + A \leftrightarrow X + \sum_i A_i$ involving light nuclei.
The validity of the law of mass action for the \emph{strong} $\pi + d \leftrightarrow \pi + n + p$ reaction during the hadronic phase was recently illustrated in Ref.~\cite{Oliinychenko:2018ugs} in a microscopic transport model calculation, keeping the deuteron yield close to the thermal model value and echoing some earlier results based on the kinetic approach~\cite{Oh:2009gx,Cho:2015exb}.
Further, the Saha equation approach implies that also the yields of hypernuclei stay virtually constant during the evolution after the chemical freeze-out and should be described by a thermal model calculation at $T = T_{\rm ch}$, in agreement with the available data on hypertriton production~\cite{Adam:2015yta}.

\begin{figure}[t]
  \centering
  \includegraphics[width=.49\textwidth]{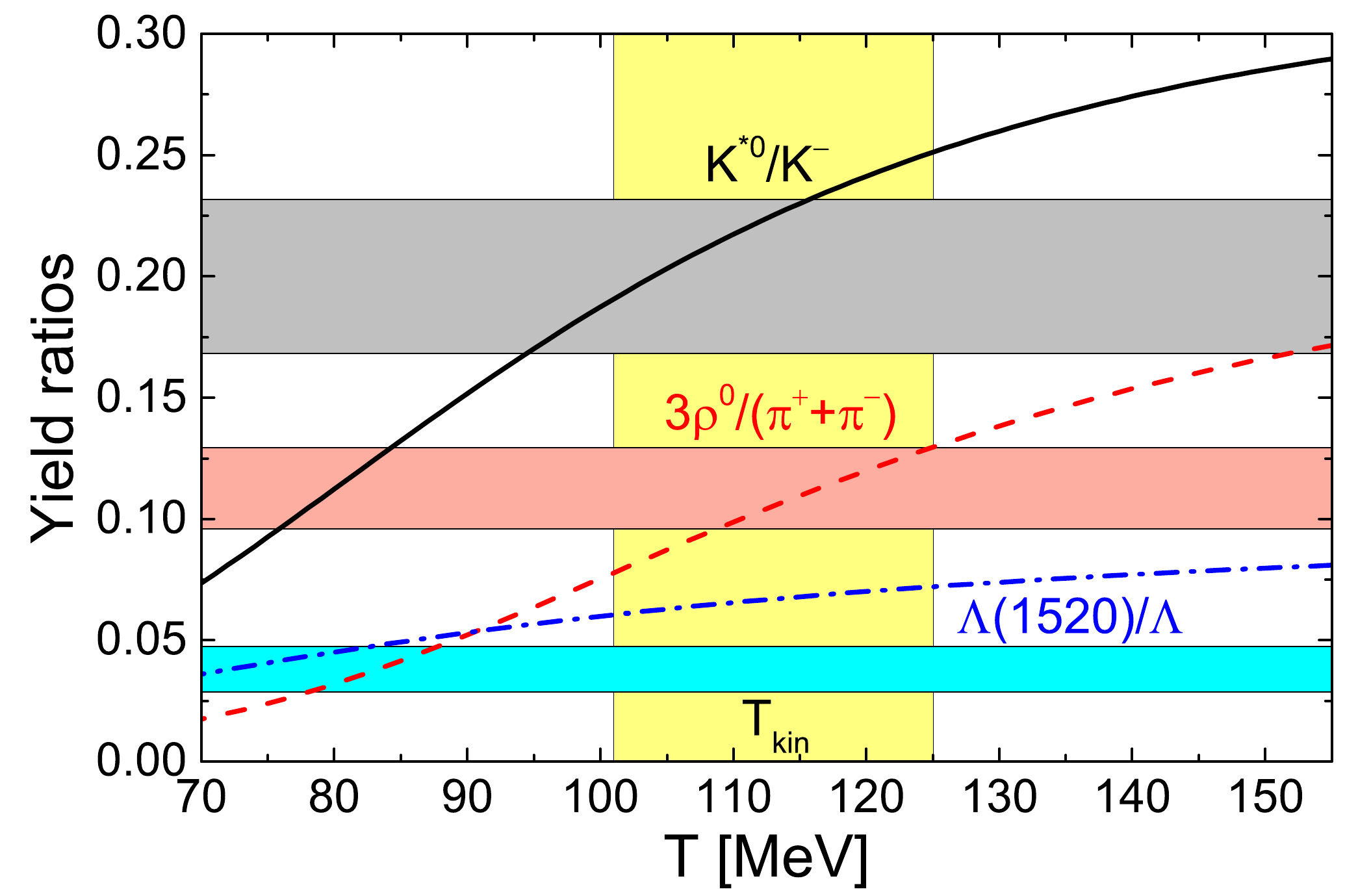}
  \caption{
   Temperature dependence of the yield ratios $\text{K}^{*0}/\text{K}^-$~(solid black line), $3\rho^0 / (\pi^- + \pi^+)$~(dashed red line), and $\Lambda(1520)/\Lambda$~(dot-dashed blue line) evaluated within the HRG in PCE at $T<T_{\rm ch}$. 
   The horizontal bands correspond to the experimental data of the ALICE collaboration for 0-20\% most central Pb--Pb collisions~\cite{Abelev:2014uua,Acharya:2018qnp,ALICE:2018ewo}.
   The vertical yellow band has the same meaning as in Fig.~\ref{fig:nuclei-vs-T}a.
  }
  \label{fig:resonances}
\end{figure}

Our approach assumes isentropic expansion after the chemical freeze-out, which essentially corresponds to an ideal hydrodynamic evolution below $T_{\rm ch}$.
The large values of the specific shear viscosity of a hadron gas reported in the literature~\cite{Demir:2008tr} may call such an assumption into question, suggesting a sizable entropy increase in the hadronic phase.
A viscous hydrodynamic evolution of a hadronic gas in PCE down to $T = 100$~MeV has been considered in Ref.~\cite{Niemi:2012ry}, where it was reported that the entropy increases by less than 1\% during that phase.
Such a result would fully justify the assumption of entropy conservation used in our work.
On the other hand, a more recent transport model study~\cite{Xu:2017akx} suggests a sizable entropy increase in the hadronic phase, mainly due to decaying resonances.
Therefore, we explore schematically the effect a possible entropy non-conservation on the light nuclei abundances in the Saha equation approach.
Namely, we consider that the total entropy at $T = 100$~MeV increases
relative to its value at $T = T_{\rm ch}$ by appropriately adjusting the r.h.s. of Eq.~\eqref{eq:pce:S}, and then calculate the nuclear abundances through the Saha equation.
We find that the entropy increase generally leads to a suppression of light nuclei yields, and the suppression is stronger for heavier nuclei.
If the relative entropy increase is mild~(within 10\%), the change in nuclear abundances is not large enough to destroy the agreement with experimental data shown in Fig.~\ref{fig:nuclei-vs-T}.
The disagreement with the data will become apparent for larger assumed values of the entropy increase, especially for $A \geq 3$ nuclei.
We leave a more rigorous treatment of the possible entropy non-conservation effect on various observables for future studies.

Accuracy of the presented results depends crucially on the validity of the PCE picture during the hadronic phase between the chemical and kinetic freeze-outs.
Existence of the hadronic phase is suggested by the lower kinetic freeze-out temperatures extracted from the blast-wave fits.
We argue that additional evidence for the validity of this picture is provided by modifications of resonance yields in the hadronic phase.
While the total yields of all hadrons stable under strong interactions are frozen during the PCE evolution, the yields of resonances are not.
This is illustrated in Fig.~\ref{fig:resonances}, where temperature dependence of the yield ratios $\text{K}^{*0}/\text{K}^-$, $3\rho^0 / (\pi^- + \pi^+)$, and $\Lambda(1520)/\Lambda$ at $T < T_{\rm ch}$ is depicted.
The resonance yields are substantially suppressed at $T \approx T_{\rm kin}$ relative to their abundances at $T = T_{\rm ch}$, as first pointed out in Ref.~\cite{Rapp:2003ar} within the PCE framework. Other considerations~\cite{Kanada-Enyo:2006dxd,Cho:2015qca} similarly predict a suppression.
The ratios $\text{K}^{*0}/\text{K}^-$ and $3\rho^0 / (\pi^- + \pi^+)$ computed at $T \approx T_{\rm kin}$ agree much better with the available experimental data for Pb--Pb collisions~\cite{Abelev:2014uua,Acharya:2018qnp} than the corresponding values at the chemical freeze-out of stable hadrons.
The $\Lambda(1520)/\Lambda$ ratio is also suppressed by about 20\%, although the data~\cite{ALICE:2018ewo} are still overestimated.
We note that this ratio might be sensitive to other effects not considered here, such as finite resonance widths~\cite{Vovchenko:2018fmh}, which could modify the $\Lambda(1520)$ yield, or extra strange baryonic resonances~\cite{Alba:2017mqu} which may increase the $\Lambda$ feeddown.

\paragraph{Summary and conclusions.}

We analyzed the production of light (anti-)(hyper-)nuclei in heavy-ion collisions at the LHC in the framework of the Saha equation, making use of the intimate and illustrative analogy between the evolution of the early universe after the Big Bang and that of ``Little Bangs'' created in the lab.
Assuming that strong disintegration and regeneration reactions involving light nuclei proceed in relative chemical equilibrium after the chemical freeze-out of hadrons, their abundances are determined through the nuclear equivalent of the Saha ionization equation, where the strong exponential dependence on the temperature typical for the standard thermal model is eliminated.
A quantitative analysis, performed using the hadron resonance gas model in partial chemical equilibrium, shows agreement with the experimental data of the ALICE collaboration on 
d, $^3$He, $^3_\Lambda$H, and $^4$He
yields for a very broad region of temperatures well below the chemical freeze-out, $T  \lesssim 155$~MeV. 
Here we focused at the LHC, but the formalism can as well be considered at lower collision energies, such as RHIC or SPS.

The presented picture, supported by the observed suppression of resonance yields in central Pb-Pb collisions at the LHC, explains the apparent agreement of thermal model predictions with the measured light (anti-)(hyper-)nuclei abundances.
The results, however, do \emph{not} imply that the nuclei are formed at the chemical freeze-out and survive the subsequent evolution.
In fact, formation of nuclei in a diluting and cooling hadronic system at virtually \emph{any} temperature below $T_{\rm ch}$ can be accommodated within the present approach, including, for instance, at the kinetic freeze-out.
The present work does not answer where and when specifically these fragile objects, hypertriton in particular, are formed. 
A quantum mechanical description of creation and decreation of (tightly) bounded states in an open thermal system
would be necessary to obtain more specific conclusions on when the light nuclei may begin to appear as bound states~(as dictated by detailed balance),
while a comparison of the fireball expansion rate relative to the reaction rates involving various nuclei is needed to establish their freeze-out.


\begin{acknowledgments}

\emph{Acknowledgments.} 
V.V. and C.G. acknowledge the participation in the ECT* workshop on ``Observables of Hadronization and the QCD Phase Diagram in the Cross-over Domain''~(Trento, Italy, October 15-19, 2018), which inspired the early ideas about the work reported here.
K.G. was supported by the Bundesministerium f\"ur Bildung und Forschung~(BMBF), grant No. 3313040033.

\end{acknowledgments}

\bibliography{SahaLHC}


\end{document}